\documentclass[aps,pra,superscriptaddress,reprint,floatfix,longbibliography,showpacs,amsfonts,amsmath,amssymb]{revtex4-1}
\usepackage{graphicx}
\usepackage{hyperref}
\usepackage{xr-hyper}
\usepackage{upgreek}
\usepackage{epstopdf}
\usepackage[note-name=, use-sort-key = false]{notes2bib}
\usepackage{color}
\usepackage{comment}

\begin{document}
\title{Indications for Lifshitz transitions in the nodal-line semimetal ZrSiTe induced by interlayer interaction}

\author{M. Krottenm\"uller}
\email{These authors contributed equally.}
\affiliation{Experimentalphysik II, University of Augsburg, 86159 Augsburg, Germany}
\author{M. V\"ost}
\email{These authors contributed equally.}
\affiliation{Chair of Chemical Physics and Materials Science, Institute of Physics, University of Augsburg, 86159 Augsburg,
Germany}
\author{N. Unglert}
\email{These authors contributed equally.}
\affiliation{Chair of Chemical Physics and Materials Science, Institute of Physics, University of Augsburg, 86159 Augsburg,
Germany}
\author{J. Ebad-Allah}
\affiliation{Experimentalphysik II, University of Augsburg, 86159 Augsburg, Germany}
\affiliation{Department of Physics, Tanta University, 31527 Tanta, Egypt}
\author{G. Eickerling}
\affiliation{Chair of Chemical Physics and Materials Science, Institute of Physics, University of Augsburg, 86159 Augsburg,
Germany}
\author{D. Volkmer}
\affiliation{Chair of Solid State and Materials Chemistry, Institute of Physics, University of Augsburg, 86159 Augsburg,
Germany}
\author{J. Hu}
\affiliation{Department of Physics, University of Arkansas, Fayetteville, AR 72701, USA}
\author{Y. L. Zhu}
\affiliation{Department of Physics, Pennsylvania State University, University Park, PA 16803, USA}
\affiliation{Department of Physics and Engineering Physics, Tulane University, New Orleans, LA 70118, USA}
\author{Z. Q. Mao}
\affiliation{Department of Physics, Pennsylvania State University, University Park, PA 16803, USA}
\affiliation{Department of Physics and Engineering Physics, Tulane University, New Orleans, LA 70118, USA}
\author{W. Scherer}
\email{wolfgang.scherer@physik.uni-augsburg.de}
\affiliation{Chair of Chemical Physics and Materials Science, Institute of Physics, University of Augsburg, 86159 Augsburg,
Germany}
\author{C. A. Kuntscher}
\email{christine.kuntscher@physik.uni-augsburg.de}
\affiliation{Experimentalphysik II, University of Augsburg, 86159 Augsburg, Germany}

\begin{abstract}
The layered material ZrSiTe is currently extensively investigated as
a nodal-line semimetal with Dirac-like band crossings
protected by nonsymmorphic symmetry close to the Fermi energy. A recent infrared
spectroscopy study on ZrSiTe under external pressure found anomalies in the
optical response, providing hints for pressure-induced phase transitions at
$\approx$4.1 and $\approx$6.5~GPa.
By pressure-dependent Raman spectroscopy and x-ray diffraction measurements combined with
electronic band structure calculations we find indications for two
pressure-induced Lifshitz transitions with major
changes in the Fermi surface topology in the absence of lattice
symmetry changes.
These electronic phase transitions can be attributed to
the enhanced interlayer interaction induced by external pressure.
Our findings demonstrate the crucial role of the interlayer distance
for the electronic properties of layered van der Waals topological
materials.
\end{abstract}
\pacs{}

\maketitle

Topological materials such as topological insulators \cite{Fu.2007},
Dirac \cite{Young.2012}, Weyl
\cite{Wan.2011,Soluyanov.2015,Armitage.2018} or line-node semimetals
\cite{Burkov.2011a,Fang.2015} are of great fundamental interest due to
their exotic nature of electronic phases, and thus heavily
investigated nowadays. They usually exhibit extraordinary material
properties, for example, high carrier mobility and unusual
magnetoresistance \cite{Neupane.2014,Lv.2016,Sankar.2017}. Topological
non-trivial phases often occur in layered materials with weak
interlayer bonding, where the single layers behave rather as isolated
two-dimensional (2D) objects, enabling the exfoliation to atomically
thin 2D crystals with numerous possible applications
\cite{Xu.2015,Schoop.2016,Deng.2016,Zhang.2019,Ma.2019}.  Since the
forces between the layers of such structures are usually weak, they
are highly compressible perpendicular to the layers, and a dimensional
crossover from 2D to 3D can be induced by external pressure.
Generally, layered materials are prone to pressure-induced phenomena,
and electronic topological transitions \cite{Lifshitz.1960} are
expected to be induced \cite{Zhu.2012,Yang.2017,Bassanezi.2018} and
were suggested to occur in layered BiTeBr \cite{Ohmura.2017}, BiTeI
\cite{Xi.2013,Ponosov.2013}, 1T-TiTe$_2$ \cite{Rajaji.2018}, and the
group V selenides and tellurides Bi$_2$Se$_3$, Bi$_2$Te$_3$,
Sb$_2$Te$_3$
\cite{Polian.2011,Gomis.2011,Vilaplana.2011a,Vilaplana.2011b}.

``Electronic transitions'' in metals were first introduced by Lifshitz
in 1960 as transitions where the topology of the Fermi surface (FS) changes
as a result of the continuous deformation under high external pressure
\cite{Lifshitz.1960}. Examples for pressure-induced alterations of the
FS topology are the conversion of an open Fermi surface,
such as a corrugated cylinder-type Fermi surface typical for layered
materials, to a closed one, or the appearance of a new split-off
region of the FS. Importantly, the changes in the Fermi
surface topology during such a so-called Lifshitz transition are not
related to a change in the lattice symmetry \cite{Lifshitz.1960}.

In this work we find indications for two Lifshitz transitions in
the layered van der Waals material ZrSiTe under external pressure,
resulting from the enhanced interlayer interaction. ZrSiTe belongs to
the family of compounds Zr$XY$ ($X$=Si, Ge, Sn and $Y$=O, S, Se, Te),
which are nodal-line semimetals
\cite{Wang.1995,Bensch.1995}. The materials Zr$XY$ exhibit largely
varying interlayer bonding \cite{Xu.2015,Ebad-Allah.2019}, with ZrSiTe
being a clear outlier towards more 2D character
\cite{Wang.1995,Bensch.1995}.  According to recent infrared reflectivity measurements
under external pressure \cite{Ebad-Allah.2019a}, ZrSiTe is highly sensitive to pressure.
In particular, several optical parameters showed anomalies in their pressure dependence at
the critical pressures $P_{c1}\approx$4.1~GPa and
$P_{c2}\approx$6.5~GPa, suggesting the occurrence of two phase
transitions of either electronic or structural type.
For the clarification of the pressure-induced effects in ZrSiTe, we carried out
Raman spectroscopy and x-ray diffraction measurements under external pressure and
density-functional-theory (DFT) electronic band structure calculations.


Zr$XY$ compounds, with $X$ being a carbon group atom and $Y$ a
chalcogen atom, have a PbFCl-type tetragonal structure with space
group $P4/nmm$ and lattice parameters $a$=3.70~\AA\ and $c$ = 9.51~\AA\ for ZrSiTe (see Fig.\ S1
in the Supplemental Material \cite{Suppl}) \cite{Wang.1995,Bensch.1995}.
The layered crystal structure of Zr$XY$ is built from quintuple layers of
[Y-Zr-X-Zr-Y] parallel to the $ab$ plane. Due to the
layered nature, the occurrence of so-called rigid-layer phonon modes
is expected \cite{Zallen.1974}, where atomic layers move against each
other along the $c$ direction.  The {\it basal-plane} Raman spectra of
ZrSi$Y$ ($Y$=S, Se, Te) and ZrGe$Y$ ($Y$=S, Se) single crystals
consist of three modes [see Fig.\ S1(b) in the Supplemental Material \cite{Suppl}] which can be assigned to the
$A^1_{1g}$, $A^2_{1g}$, and $B_{1g}$ modes.  The atomic displacements
of these three Raman modes consist of mostly out-of-plane motions
along the $c$ direction
\cite{Zhou.2017,Singha.2018,Sorb.2013}. Hereby, the modes $A^1_{1g}$
and $A^2_{1g}$ can be assigned to motions of Zr and $Y$ atoms and the
mode $B_{1g}$ to motions of $X$ atoms.

\begin{figure}[t]
\includegraphics[width=0.45\textwidth]{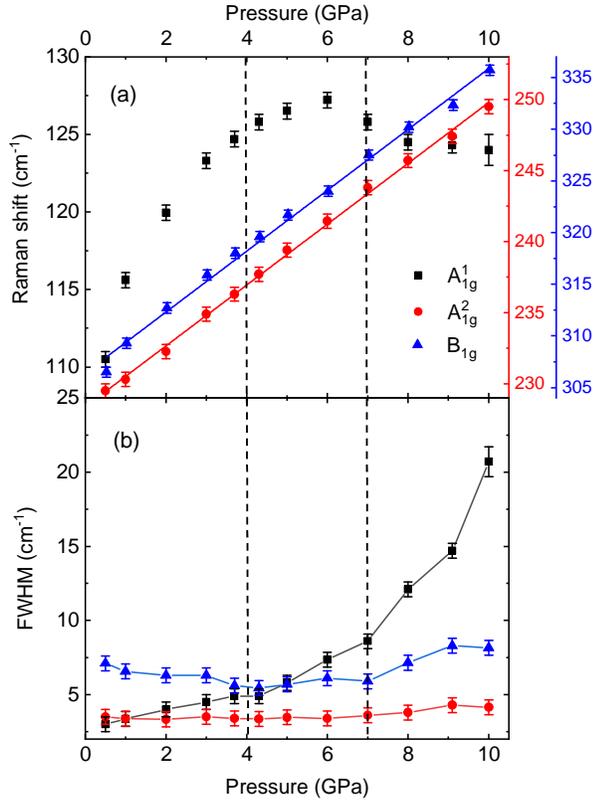}
\caption{(a) Frequencies of the {\it basal-plane} Raman modes
  $A^1_{1g}$, $A^2_{1g}$, and $B_{1g}$ in ZrSiTe as a function of
  pressure. The red and blue lines are linear fits of the pressure
  evolution of the $A^2_{1g}$ and $B_{1g}$ mode frequencies,
  respectively. (b) FWHM of the Raman modes $A^1_{1g}$, $A^2_{1g}$,
  and $B_{1g}$ in ZrSiTe as a function of pressure. The two vertical
  dashed lines indicate the two critical pressures $\sim$4 and
  $\sim$7~GPa.}\label{figure3}
\end{figure}

\begin{figure}[t]
\includegraphics[width=0.4\textwidth]{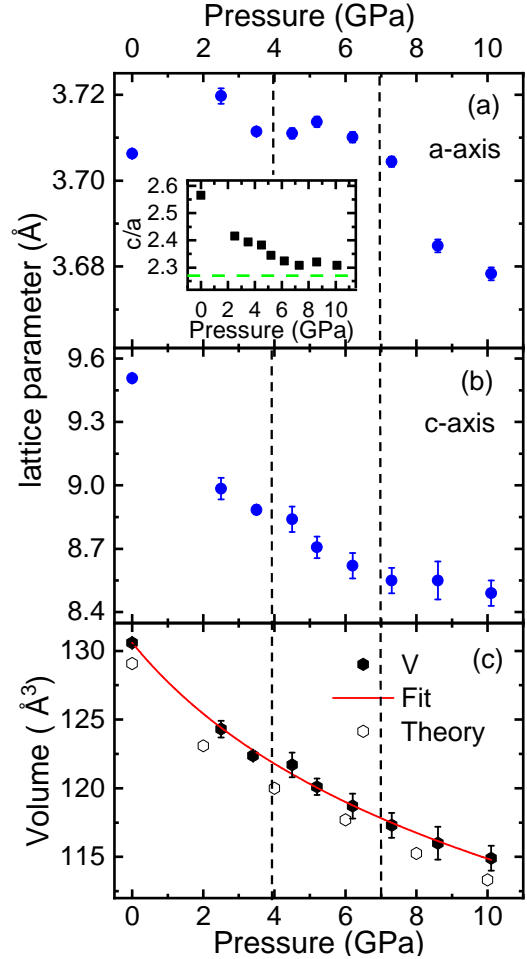}
\caption{(a), (b): Pressure dependence of the lattice parameters $a$
  and $c$ of ZrSiTe. Inset: Pressure dependence of the $c$/$a$
  ratio. The horizontal, dashed line marks the $c$/$a$ ratio of ZrSiS at ambient
  pressure \cite{Singha.2018,Gu.2019}.  (c) Volume $V$ of the unit cell as a function
  of pressure together with the calculated values. The solid
  line is a fit of the experimental data with a second-order Murnaghan EOS as defined in the
  text. The two vertical
  dashed lines indicate the two critical pressures $\sim$4 and
  $\sim$7~GPa.}\label{figure-parameters}
\end{figure}

\begin{figure*}
\includegraphics[width=1\textwidth]{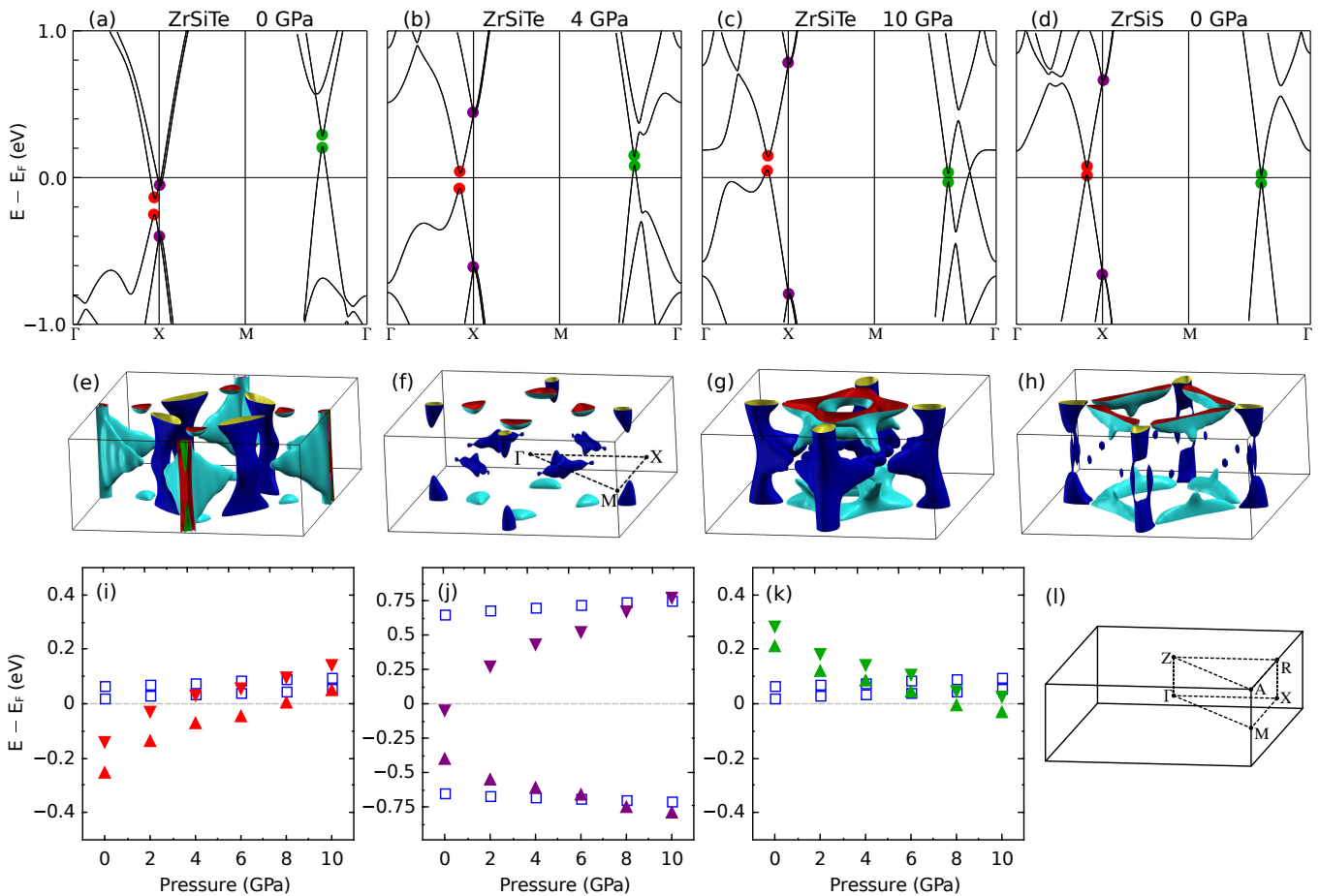}
\caption{(a) - (c) Electronic band structure of ZrSiTe at selected
  pressures, with the Dirac crossings marked by colored points. (d) Electronic band structure of ZrSiS at ambient
  pressure.  (e) - (g) Fermi surface of ZrSiTe at selected
  pressures. (h) Fermi surface of ZrSiS at ambient pressure. (i) -
  (k) Energy position of Dirac crossings in the electronic band
  structure of ZrSiTe [triangles, color coding according to the points in (a) - (c)] and ZrSiS (squares) as a function of
  pressure. (l) High-symmetry points in the tetragonal BZ.}
\label{fig:theory}
\end{figure*}

According to the pressure-dependent Raman spectra of ZrSiTe (see
Fig. S2 in the Supplemental Material \cite{Suppl}) all three phonon modes
$A^1_{1g}$, $A^2_{1g}$, and $B_{1g}$ harden with increasing pressure
up to 6~GPa and are observed up to the highest measured
pressure (10~GPa).
Importantly, there are no signs for pressure-induced structural phase transitions.
The pressure dependence of the frequencies was extracted by
Lorentz fitting and is depicted in Fig.\ \ref{figure3}(a). The
pressure dependence of the modes $A^2_{1g}$ and $B_{1g}$ is linear
with the pressure coefficients 2.1 and 2.9~cm$^{-1}$/GPa,
respectively. In contrast, we observe a
pronounced nonlinearity of the pressure-induced shift of the
$A^1_{1g}$ mode: Up to $\sim$4~GPa this mode shows a strong hardening
with increasing pressure, which saturates between 4 and
6~GPa. Remarkably, above $\sim$7~GPa the $A^1_{1g}$ mode softens with
increasing pressure. Also the pressure
behavior of the full-width at half-maximum (FWHM) of the $A^1_{1g}$
mode is distinct [see Fig.\
\ref{figure3}(b)]: Whereas the FWHM of the $A^2_{1g}$ and $B_{1g}$
modes shows no clear pressure dependence, the FWHM of $A^1_{1g}$
increases linearly up to $\sim$7~GPa, with
an anomaly at $\sim$4~GPa above which the linear pressure coefficient
increases. Above $\sim$7~GPa the FWHM of $A^1_{1g}$ increases
strongly \bibnote{Preliminary DFT calculations on the pressure dependency
of the Raman phonon mode frequencies and linewidths hint at possible anharmonicity
effect with increasing pressure. Details will be presented in a subsequent paper.}.

In analogy to ZrSiS and PbFCl, the $A^1_{1g}$ mode is ascribed to the
relative motion between two weakly bonded Zr-$Y$ units
\cite{Singha.2018,Sorb.2013}. Due to the in-phase motion of the $Y$
layer and the adjacent Zr layer (see Fig.\ S1(a) in the Supplemental
Material \cite{Suppl} for illustration) it can be considered as a rigid-layer mode,
which is very sensitive to changes in interlayer interactions induced,
e.g., by external pressure. In particular, the frequencies of
rigid-layer modes can be employed to monitor the changes in the
interlayer bonding, as demonstrated for layered chalcogenide crystals
\cite{Zallen.1974}.  Accordingly, the observed strong pressure-induced
hardening of the rigid-layer $A^1_{1g}$ mode up to 6~GPa signals a strong enhancement
of the interlayer interaction in ZrSiTe during pressure application.
It is interesting to note that, previously, pressure-induced anomalies in the frequency
and FWHM of Raman modes were interpreted in terms of electronic phase
transitions (EPTs) in several cases
\cite{Gomis.2011,Vilaplana.2011a,Vilaplana.2011b,Bera.2013,Ponosov.2013,Rajaji.2018}. In
particular, the FWHM can be strongly affected by changes in the
electron-phonon coupling, which can serve as an indirect signature for
an EPT \cite{Rajaji.2018}.

The pressure-induced changes in the crystal structure of ZrSiTe were
investigated by single-crystal x-ray diffraction measurements.  The
pressure dependence of the lattice parameters $a$ and $c$
and of the unit cell volume $V$ is depicted in Fig.~\ref{figure-parameters}.
The volume $V$ decreases monotonically with increasing
pressure $P$ and can be described by a second-order
Murnaghan equation of state (EOS) \cite{Murnaghan.1944}:%
\begin{equation}
V(P)=V_{0}\cdot\lbrack(B_{0}^{\prime}/B_{0})\cdot P+1]^{-1/{B_{0}^{\prime}}}%
\end{equation}%
where $B_{0}$ is the bulk modulus, $B_{0}^{\prime}$ its pressure derivative, and $V_0$
the volume, all at $P$=0~GPa. The so-obtained bulk modulus
amounts to $B_{0}$=40.9$\pm$2.5~GPa and its derivative to
$B_{0}^{\prime}$=9.2$\pm$1.0. The value of $B_{0}^{\prime}$ of
ZrSiTe is enhanced compared to the value $B_{0}^{\prime}$=4 typical
for 3D materials with isotropic elastic properties,
which signals the layered character of ZrSiTe. For comparison, the
corresponding values of $B_{0}$ and $B_{0}^{\prime}$ of graphite
amount to 33.8~GPa and 8.9, resp., \cite{Hanfland.1989} and for the
more 3D sister compound ZrSiS 141$\pm$4.5~GPa and 5.1$\pm$0.5, resp.\ \cite{Singha.2018}.

The lattice parameters $a$ and $c$ show a distinct pressure dependence
for the pressure range below and above $\sim$7~GPa (see Fig.\ \ref{figure-parameters}).
Whereas parameter $a$ is hardly affected by pressure (please note the scale),
parameter $c$ monotonically decreases up to 7~GPa and is pressure independent above 7~GPa.
Obviously, the $c$ direction perpendicular
to the layers shows the highest compressibility and is most affected
at low pressures. This finding is also illustrated by the $c$/$a$
ratio, which basically follows the pressure dependence of the parameter $c$
[see inset of Fig.\ \ref{figure-parameters}(a)].
It is important to note that, despite these changes, we
can exclude the occurrence of a crystal symmetry change for pressures
up to 10~GPa (see Supplemental Material \cite{Suppl} for details).

For an interpretation of the experimental results,
we performed first-principles DFT electronic structure calculations.
The electronic band structure of ZrSiTe for selected
pressures together with the corresponding FS is shown in
Figs.\ \ref{fig:theory}(a)-(c) and (e)-(g), respectively. In agreement
with earlier reports \cite{Topp.2016,Hosen.2017,Ebad-Allah.2019}, the
ambient-pressure electronic band structure of ZrSiTe contains a nodal line,
which is gapped due to spin-orbit coupling. Additional Dirac-like band
crossings, which are protected by nonsymmorphic symmetry against
gapping (called nonsymmorphic Dirac crossings in the following) appear
close to the Fermi energy $E_F$ at the $X$ and $R$ point of the
Brillouin zone (BZ).  The nodal line close to $E_F$, which is found in all
Zr$XY$ compounds, is formed by Si $sp_xp_y$-Zr$d$ hybrid orbitals
\cite{Ebad-Allah.2019,Bensch.1995}. In fact, the two linearly crossing
(touching) bands in Zr$XY$ are located along a surface in the BZ, forming an effective nodal plane
\cite{Ebad-Allah.2019}.  The FS of ZrSiTe has a
diamond-rod-shape with additional four, rather flat pillars and four
electron pockets [see Fig.\ \ref{fig:theory}(e)].

From our pressure-dependent calculations it is obvious that the
electronic band structure of ZrSiTe is highly sensitive to
external pressure. With increasing pressure, the diamond-shaped
FS shrinks to be small, and the pillars are disrupted as well. Overall, the
FS is drastically reduced up to $\sim$4~GPa (see Fig.\
\ref{fig:theory} and the Supplemental Material \cite{Suppl}), in agreement with the
reported decrease of the plasma frequency
\cite{Ebad-Allah.2019a}. Above 4~GPa this trend is reversed, i.e., the
FS enlarges with increasing pressure, and above $\sim$8~GPa
some parts of the FS are connected, forming corrugated
cylinders with side arms and a diamond-shaped component. This is in
full agreement with the behavior of the plasma frequency, showing an
increase above 4~GPa followed by a plateau-like saturation above
$\sim$7~GPa, reaching a value close to the ambient-pressure one
\cite{Ebad-Allah.2019a}. From these drastic pressure-induced changes
of the FS, we infer that ZrSiTe undergoes two Lifshitz
transitions in the pressure range 0 - 10~GPa.

The pressure-dependent energy position of the Dirac crossings in the electronic
band structure of ZrSiTe and ZrSiS are depicted in Figs.\ \ref{fig:theory}(i)-(k).
With increasing pressure
the Dirac crossings forming the nodal line are shifted towards $E_F$
and even cross $E_F$ above $\sim$4~GPa.  The nonsymmorphic Dirac
crossings are strongly pushed away from the Fermi level. At the
highest studied pressure (10~GPa) the electronic band structure and
FS of ZrSiTe has strong similarities with the sister
compound ZrSiS at ambient pressure [see Figs.\ \ref{fig:theory}(d) and
(h)], with the nonsymmorphic Dirac crossings at energies $\pm$0.5~eV away
from E$_F$ [Fig.\ \ref{fig:theory}(j)]. Hence, high-pressure ZrSiTe
provides the opportunity to extract the properties of the nodal-line state without the influence of the
nonsymmorphic Dirac state, similar to ZrSiS but with larger spin-orbit coupling.
Together with the fact that the $c$/$a$ ratio of ZrSiTe at
10~GPa is close to that of ambient-pressure ZrSiS \cite{Singha.2018,Gu.2019} [see Fig.\
\ref{figure-parameters}(a)], these findings further stress the
importance of the interlayer interaction for the electronic properties
of the Zr$XY$ compound family.

According to our results, we propose the following
scenario for ZrSiTe under pressure: For
0$\lesssim$$P$$\lesssim$4~GPa a pressure-induced dimensional crossover
from layered to more 3D is induced within the parent tetragonal phase,
as indicated by the drastic decrease in the $c$/$a$ ratio and the strong
hardening of the $A^1_{1g}$ mode, concomitant with its intensity increase.
At $\sim$4~GPa a Lifshitz transition occurs with drastic changes in the
electronic band structure and a shrinkage of the FS,
leading to an intermediate phase for pressures
4$\lesssim$$P$$\lesssim$7~GPa.
At $\sim$7~GPa another Lifshitz
transition with an enlargement of the FS occurs, and the electronic band structure of
ZrSiTe becomes similar to that of ambient-pressure ZrSiS.

Recently, a temperature-induced Lifshitz transition was observed in
the layered topological material ZrTe$_5$ \cite{Zhang.2017,Xu.2018}. It was speculated that
this transition is induced by the variation of the
interlayer interaction with temperature.
Our results for the van der Waals material ZrSiTe show that
layered topological materials are prone to Lifshitz transitions driven
by the pressure-induced enhancement of the interlayer interaction.

In summary, by Raman spectroscopy we observed anomalies in the pressure dependence of the
frequency and line-width of the rigid-layer phonon mode in ZrSiTe, in the absence of any lattice symmetry
change according to pressure-dependent x-ray diffraction results.
The pressure behavior of the Raman mode can be explained
by the drastic decrease of the $c$/$a$ lattice parameter ratio
concomitant with an enhanced interlayer interaction.
LDA band structure calculations indicate the occurrence of two
Lifshitz transitions at $\sim$4~GPa and $\sim$7~GPa with drastic
changes in the Fermi surface topology. The Lifshitz transitions can be attributed to the
pressure-induced enhancement of the interlayer interaction.
Our results demonstrate the crucial role of the interlayer distance in
determining the electronic structure of the layered, nodal-line semimetal ZrSiTe,
and we propose that this finding holds for layered van der Waals
topological materials in general.

\begin{acknowledgments}
  We thank Hana Bunzen for technical support.  C.A.K. acknowledges
  financial support from the Deutsche Forschungsgemeinschaft (DFG),
  Germany, through grant no.\ KU 1432/13-1. The sample synthesis and
  characterization efforts were supported by the US Department of
  Energy under grant DE-SC0019068.
\end{acknowledgments}


\begin{thebibliography}{72}
\bibitem [{1()}]{1}%
  \BibitemOpen
  \href@noop {} {}\bibinfo {note} {Preliminary DFT calculations on the pressure
  dependency of the Raman phonon mode frequencies and linewidths hint at
  possible anharmonicity effect with increasing pressure. Details will be
  presented in a subsequent paper.}\BibitemShut {Stop}%
\end{thebibliography}
\end{document}